\documentclass[aps,pra,12pt,tightenlines]{revtex4}

\begin{document}

\title{Reading QBism: A Reply to Nauenberg}

\author{Christopher A. Fuchs$^{1,2}$}
\author{N.~David Mermin$^{1,3}$}
\author{R\"udiger Schack$^{1,4}$}

\affiliation{$^1$Stellenbosch Institute for Advanced Study (STIAS), Wallenberg Research Centre
at Stellenbosch University, Marais Street, Stellenbosch 7600, South Africa\\
$^2$Department of Physics, University of Massachusetts Boston,
100 Morrissey Boulevard, Boston, Massachusetts 02125, U.S.A.\\
$^3$Laboratory of Atomic and Solid State Physics, Cornell University, Ithaca NY 14853, U.S.A.\\
$^4$Department of Mathematics, Royal Holloway University of
London, Egham, Surrey TW20$\;$0EX, UK}

\date{4 February 2015}

\begin{abstract}
This is a reply to Michael Nauenberg's arXiv:1502.00123, to be published in
the American Journal of Physics, in which he comments
critically on our paper ``An introduction to QBism with an application to the
locality of quantum mechanics'', Am.\ J. Phys. 82, 749--754 (2014) and arXiv:1311.5253. 
\end{abstract}

\maketitle

Although the meaning of the quantum state (``$\psi$") seems evident to Michael
Nauenberg, among the physicists and philosophers interested in quantum
foundations, there continues to be broad and irreconcilable disagreement even
after 90 years.  Can these 90 years of widespread confusion have something to
do with almost all physicists taking for granted the frequentist interpretation
of probability?  It seems a possibility worth considering.

The subjective view of probability, dating back to Laplace, and eloquently
advocated by B.~de Finetti, L.~J.~Savage, R. Jeffrey, and many others, is
introduced on our first page.  QBism explores what such an understanding of
probability implies for the interpretation of quantum mechanics.  Our assertion
(``without any justification'') further along the page, beginning ``since
probabilities are the personal judgements of an agent$\ldots$'', refers back to
the {\it premise\/} of subjective probability, whose implications we are about
to examine.

An immediate consequence is that the quantum state an agent assigns to a system
depends on what the agent believes about that system.  This is not unique to
QBism.  That different agents can assign different states is the point of
Wigner's famous parable about his friend.  The probabilities determined by the
Born rule are contingent on the state assignment, whether one understands those
probabilities from a subjectivist or frequentist perspective.  Any experiment
that validates or invalidates the standard link between the Born rule and the
state assignment for a frequentist does so for a subjectivist too.

QBism is not about the validity of quantum mechanics, but about how to
understand the basic concepts that appear in the theory: states, probabilities,
measurements, outcomes.  Is the state of a system an objective fact about that
system (as Nauenberg seems to believe) or is it a judgment made by a particular
agent on the basis of her prior experience of that system (the QBist view)?  Is
the outcome of a measurement a permanent record of an experiment made ``by a
macroscopic and time irreversible process", or is it the personal experience
induced in an agent by the response of her external world to any action she
takes upon it?

Frequencies are indeed ubiquitous in physics.  But the subjective theory of
probability distinguishes between frequencies and probabilities.  Frequencies
are data; probabilities are personal degrees of belief.  Frequencies can be
assigned probabilities.  And probabilities can be refined in the light of
subsequently measured frequencies.  A famous theorem of de Finetti relates the
two.

Nauenberg misses a central point of QBism in his criticism of our discussion of
nonlocality.  Events are deductions an agent makes to account for her
experience.  The correlations each agent extracts from quantum mechanics are
not between disembodied ``events", but between the experiences (outcomes of her
actions on the world) from which she constructs such events.

Nauenberg concludes that ``Contrary to Fuchs {\it et al.\/}, quantum theory
deals with the objective world as directly as does classical mechanics''.
Setting aside our doubts about his ``essential difference" between classical
and quantum mechanics, we only remark that the QBist understanding of science
applies to classical as well as quantum physics.  Unlike QBism, CBism is not
needed to resolve a scandalous incoherence at the foundations of the subject,
but it does succeed in clearing up at least one long-standing
puzzle [1].

QBism is a genuinely novel way of thinking about the function of science.  It
raises subtle questions about the nature of science, the nature of human
experience, and the relation of scientists to each other and to the world they
are attempting to understand.  We welcome criticism, but urge critics to pay
some attention to what we are saying.

\vspace{10mm}

\noindent
[1] N. David Mermin, ``QBism puts the scientist back into
  science'',   Nature {\bf 507}, 421--423 (2014).

\end{document}